\documentclass[a4paper]{article}

\begin{document}

\title{Photon filters in a microwave cavity}
\author{J. Larson and S. Stenholm  \\
Physics Department,\\
Royal Institute of Technology,\\
SCFAB, Roslagstullsbacken 21, S-10691 Stockholm, Sweden\\}
\date{\today}
\maketitle

\begin{abstract}
In an earlier paper we have concluded that time-dependent
parameters in atom-mode interaction can be utilized to modify the
quantum field in a cavity. When an atom shoots through the cavity
field, it is expected to experience a trigonometric time
dependence of its coupling constant. We investigate the
possibilities this offers to modify the field. As a point of
comparison we use the solvable Rosen-Zener model, which has
parameter dependencies roughly similar to the ones expected in a
real cavity.  We do confirm that by repeatedly sending atoms
through the cavity, we can obtain filters on the photon states.
Highly non-classical states can be obtained. We find that the
Rosen-Zener model is more sensitive to the detuning than the case
of a trigonometric coupling.
\end{abstract}

\section{Introduction}
The Jaynes-Cummings model (\cite{jaynes-cummings} and
\cite{shore-knight}) has been widely used to model the bahaviour
of atoms in high-quality cavities. Usually it has been used with
constant parameters, and the possibility to obtain analytic
results for the time evolution has made it a useful tool in
analyzing both the theoretical and experimental aspects of cavity
QED. In our earlier works (\cite{jonas} and \cite{fredrik}) we
have considered the possibility to tune the cavity parameters
during the time of interactions with the individual atoms. In
\cite{jonas}, we utilized the fact that the model reduces to sets
of uncoupled two-level systems in order to manipulate the quantum
state of the cavity. Such modifications of the parameters must
take place slowly enough to allow the cavity to follow
adiabtically retaining the identity of its individual modes. By
introducing exactly solvable models, we were able to explore the
possibilities offered by such systems. In particular, we
considered the time evolution during the interaction followed by a
projective measurement on the emerging atom as a tool to modify
the shape of the photon distribution in the cavity; the atoms
acted as filters on the photon state.

The Jaynes-Cummings model has been used extensively to model the
physics in realistic cavity QED experiments based on microwave
radiation \cite{walther}. Schemes for preparing various states of
the microwave cavity field have been proposed and realized, see
e.g. \cite{field-prep}.  In such cavities the shape of the
electromagnetic modes is mainly given by standing waves
approximately described by trigonometric functions. The atom is
supposed to enter the cavity at a node and exit at an opposite
node. An atom thus travelling through the cavity sees this mode
shape as a time dependence of the coupling constant. The model
cannot be solved analytically for this case, except when the
detuning is exactly zero. Such a solution was first given by
Schlicher \cite{schlicher} in the present context. This work,
however, was mainly concerned with the effect of the field on the
atom not exploring the effect the atomic observation have on the
field. Some other papers \cite{tidmod} and \cite{joshi}  also
considers the effects of the time dependence induced by the mode
structure traversed by the atom. The present use of the time
dependence to shape the quantum state of the field is not
considered in these papers.

We also, however, want to consider the exactly solvable Rosen-Zener model, which has a time
dependence of the coupling function which approximates the mode shape in a
cavity. Here as in the actual cavity, the detuning is taken
to be constant. This neglects possible Stark shifts caused by the
electromagnetic field. When an atom travels through the standing wave
of a real cavity,
it thus experiences a
time dependent coupling which does not greatly differ from that in the
model.

The Rosen-Zener model was not investigated in our previous work
\cite{jonas}, but here we use it as a point of comparison for the
more realistic trigonometric dependence. We find that the exact
shape of the coupling function is not crucial, but also that the
Rosen-Zener model is much more sensitive to the detuning than the
cavity mode model.

The section \ref{teori}
summarises the results from paper \cite{jonas} on the time-dependent Jaynes-Cummings
model and section \ref{r-z} applies these results to the Rosen-Zener model. In
section \ref{microw} we present the details of our more realistic description of a
microwave mode traversed by a sequence of single atoms. Section \ref{numeri}
compares the numerically obtained results for the cavity mode with those of
the Rosen-Zener and shows how efficient the type of interaction is, if we
want to achieve a narrow photon distribution. Section \ref{con}
presents some general observations on the
results obtained.

\section{Time-dependent Jaynes-Cummings model}
\subsection{Manipulating the cavity field}\label{teori}
The Jaynes-Cummings model describes the atom
 by a two-level state $|\pm\rangle$ and the field consists of
a single mode. Further, the cavity losses and spontaneous emission of
 the atomic levels are
neglected, which means that characteristic time-scales in the
 interacting system are
much shorter than the atomic and the cavity loss time. The Jaynes-Cummings Hamiltonian
is (with $\hbar=1$)
\begin{equation}
H=\Omega b^{\dagger}b+\frac{\omega}{2}\sigma_3+g\left(b^{\dagger
}\sigma^-+b\sigma^+\right),
\end{equation}
where the $\sigma$:s are the ordinary Pauli matrices.

The operator $N=b^{\dagger}b+\frac{1}{2}\sigma_3$ commutes
with $H$ and is therefore a constant of motion. We consider
the Hamiltonian $\tilde{H}=H-\Omega N$ and the dynamics of $N$ can be
added separately afterwards. In the basis $\Big\{
  |n,\pm\rangle\Big\}$,
the transformed Hamiltonian $\tilde{H}$ is in
block-diagonal form, and the Schr\"odinger equation can be solved within
each block
\begin{equation}
i\frac d{dt}\left[
\begin{array}{l}
a_{+}(n) \\
\\
a_{-}(n)
\end{array}
\right] =\left[
\begin{array}{lll}
\displaystyle{\frac{\Delta \omega }{2}} &  & g\sqrt{n} \\
&  &  \\
g\sqrt{n} &  & \displaystyle{-\frac{\Delta \omega }{2}}
\end{array}
\right] \left[
\begin{array}{l}
a_{+}(n) \\
\\
a_{-}(n)
\end{array}
\right] ,  \label{a6}
\end{equation}
with the detuning $\Delta\omega=\omega-\Omega$. As mentioned in the
introduction, here the detuning is assumed to be time-independent while the coupling constant $g$ depends on time.

The state of the whole system can be written as
\begin{equation}
|\Psi\rangle=c_0a_-(0)|0,-\rangle+\sum_{n=1}^{\infty}c_n\Big[a_+(n)|n-1,+\rangle+a_-(n)|n,-\rangle\Big],
\end{equation}
where the initial state, determined by $c_n$ and $a_{\pm}^0(n)$, is
assumed to be known. The state after the interaction is given by
$a_{\pm}^{\infty}(n)$.

If the atom is
found in one of the states $|\pm\rangle$ after the interaction, the
corresponding photon distribution for the cavity mode is then (up to a
normalization constant)
\begin{equation}\label{Ppm}
\begin{array}{l}
P_n^+=|a_+^{\infty}(n+1)|^2|c_{n+1}|^2 \\ \\
P_n^-=|a_-^{\infty}(n)|^2|c_n|^2.
\end{array}
\end{equation}

If we only consider the initial conditions
\begin{equation}\label{in}
|a_-^0|=1,\,\,\,\,\,\,\,a_+^0=0,
\end{equation}
or vice versa, no interference terms need to be considered, see
\cite{jonas}. From (\ref{Ppm}) it is clear that the initial field
distribution $|c_n|^2$ is modified by $|a_{\pm}^{\infty}(n)|^2$
after the measurement of the atomic state. The shape of these
``filter functions'' is crucial for what final state the field
will be found in.

For zero detuning the Schr\"odinger equation is analytically solvable \cite{schlicher}. With the
initial condition (\ref{in}), the filter functions will be oscillating
with $n$ according to
\begin{equation}\label{zerodet}
\begin{array}{lll}
|a_-^{\infty}(n)|^2 & = & \cos^2\left(\sqrt{n}A\right) \\ \\
|a_+^{\infty}(n)|^2 & = & \sin^2\left(\sqrt{n}A\right),
\end{array}
\end{equation}
with the area of the coupling given by
\begin{equation}
A=\int_{-\infty}^{\infty}g(t)dt.
\end{equation}
It follows
that the shape of $g(t)$ is not important in this case, only the
area. We note that no transitions occur if $A$ vanishes. This happens in
the trivial situation when the coupling $g(t)$ is identically equal to
zero, but it is also possible if the coupling changes sign.
For a small
detuning $\Delta\omega$ we expect solutions similar to (\ref{zerodet}), but with some
corrections of the amplitudes. When the detuning is increased further, the solutions may not
necessarily
have the oscillating form  in (\ref{zerodet}).

After the passage of one atom, the process can be repeated and, even
after several atoms, the state of the cavity mode is fully determined. For
example, if $m$ atoms are injected and all the atoms are found in
their lower state $|-\rangle$ after the interactions, the field
distribution will be
\begin{equation}\label{mner}
P_n^-(m)\propto |a_-^{\infty}(n)|^{2m}|c_{n}|^2.
\end{equation}
This method can then be used to create, for example non-classical
states of the field, provided that $|a_-^{\infty}(n)|^{2m}$ has a
sharp peak where the initial photon distribution differs from
zero. Note, however, that if atoms are detected in their upper
level, the photon distribution is multiplied by
$|a_+^{\infty}(n)|^2$ and the asymmetry in (\ref{Ppm}) makes the
analysis more complicated. This situation will be discussed
separately.

\subsection{The Rosen-Zener model}\label{r-z}
To demonstrate how the method described above can be
used to manipulate the state of the field, we look at the
Rosen-Zener model which is analytically solvable for the filter
functions. In the model the coupling is given by
\begin{equation}
g(t)=g_0\,\mathrm{sech}\!\left(\frac{t}{T}\right).
\end{equation}
and the detuning $\Delta\omega$ is constant. With the initial condition (\ref{in}), the filter functions becomes
\begin{equation}\label{rz}
\begin{array}{lll}
\mid a_{+}^\infty (n)\mid ^2 & = & \sin ^2\left( \pi
    Tg_0\sqrt{n}\right)\,\mathrm{sech}^2\!\left(\pi T\Delta \omega /2 \right) \\
&  &  \\
\mid a_{-}^\infty (n)\mid ^2 & = & 1-\mid a_{+}^\infty (n)\mid ^2.
\end{array}
\label{a20}
\end{equation}
In the non-adiabatic limit ($\Delta\omega T\approx0$), the hyperbolic
secant is equal to one, and the filter function for a lower level
projection is a simple cos-function squared just as the solutions (\ref{zerodet}). This gives a filter
around its maxima
\begin{equation}
n_M=\frac{k^2}{T^2g_0^2},\,\,\,\,\,\,k=0,1,2,...
\end{equation}
and the width of the filter function is
\begin{equation}
\Delta n_a=\frac{\sqrt{n_M}}{Tg_0}.
\end{equation}
Thus for an initial field distribution centered around one maximum
$n_M$ and with a width $\Delta n_p$ smaller than $\Delta n_a$, a
measurement of a lower level atom will decrease the width of the
photon distribution. Note that for an initial coherent state we
have $\Delta n_p\sim\sqrt{\bar{n}}$ and the filtering effect is
independent of which maximum the field is centered around. From
the solutions (\ref{rz}) we also note that as small a detuning as
possible is preferable, if we wish to sharpen the photon
distribution.

\section{The microwave cavity}
\subsection{The model}\label{microw}
We consider a microwave cavity with one single mode such that the
number of half wavelengths of the mode is $l$. Note that $l$ is not
too big, since we assume to be in the microwave regime. If we now let
an atom, moving with velocity $v$, pass through the cavity along the
standing wave, the coupling between the mode and the atom is
\begin{equation}\label{microcoup}
g(t)=\left\{\begin{array}{lll} g_0\cos(kv\,lt), &
    -\frac{\pi}{2kv}\leq t\leq\frac{\pi}{2kv} & \mathrm{and}\,\,\, \it{l}\mathrm{=1,3,...} \\ \\ g_0\sin(kv\,lt), &
    -\frac{\pi}{2kv}\leq t\leq\frac{\pi}{2kv} & \mathrm{and}\,\,\, \it{l}\mathrm{=2,4,...} \\ \\
0 & t<-\frac{\pi}{2kv} & \mathrm{or}\,\,\,t>\frac{\pi}{2kv}.\end{array}\right.
\end{equation}
The detuning is still assumed to be time-independent. No analytic
solution of the Schr\"odinger equation is known with the
coupling given by (\ref{microcoup}) and a non-zero detuning. However, if the detuning is small,
the solution is expected to be approximately given by (\ref{zerodet}). Also, in the
Rosen-Zener model we have seen that the smallest detuning
possible will enhance the
filtering effect. In this limit, we note that if $l$ is an even number,
no transition will take place, since the area $A$ vanishes. However if
$l$ is odd, then $A=2g_0/kvl$ is non-zero and we get oscillating
solutions as in (\ref{zerodet}). Thus it follows, if $l$ is odd,
that it is possible to change the state of the field by letting
atoms pass through the cavity and detect the states of the emerging
atoms. Just as in the Rosen-Zener model, non-classical states of the
field can be achived.

\subsection{Numerical results}\label{numeri}
%\begin{center}
%\vspace{1cm} Insert figure~\ref{fig1} about here \vspace{1cm}
%\end{center}

The Schr\"odinger equation (\ref{a6}) with the coupling given by
(\ref{microcoup}) has been solved numerically using the ordinary
Runge-Kutta method. The filter functions obtained are compared with
the ones in the Rosen-Zener model. In order to compare them, the
pulse-area $A$ must be choosen equal in the two models,
which means that $T=\pi/2kvl$. Further, since the area $A$ is
proportional to $l^{-1}$, we rescale it by multiplying $g_0$ with $l$. This
just changes the oscillation frequency and is done in order that solutions
with different $l$:s may be compared easily.

The  function $|a_-^{\infty}(n)|^2$ for the microwave case and the
Rosen-Zener model is plotted in figure 1 for the dimensionless
parameters $g_0=5$, $T=0.1$, $\Delta\omega=0.5$ and $l=1,2,3$. The
figure confirms that, as long as $\Delta\omega$ is small, the form
of $g(t)$ is not important. We clearly see that all three curves
with non-vanishing coupling area $A$ coincide closely. A
difference between the curves can only be seen at the maxima and
minima where the two models differ slightly; however the $l=1$ and
$l=3$ curves are identical. The curve with $l=2$ has $A=0$ and no
effect is expected to arise as is confirmed in the plot.
%\begin{center}
%\vspace{1cm} Insert figure~\ref{fig2} about here \vspace{1cm}
%\end{center}

For larger values of $\Delta\omega$, the shape of the coupling
$g(t)$ becomes more important. In the Rosen-Zener model we still
have an oscillating function, but the minima of
$|a_-^{\infty}(n)|^2$ will now differ from zero since
$\mathrm{sech}^2(\pi T\Delta\omega)<1$. It is interesting to look
for this behavior in the microwave model. Figure 2 is again
showing $|a_-^{\infty}(n)|^2$ for the two models with the same
parameters as in figure 1, except for the detuning
$\Delta\omega=5$. The filter functions for the microwave model
seem to depend more weakly on $\Delta\omega$ than in the
Rosen-Zener model. We also note that the value of the minima is
not constant, but is approaching zero for larger $n$:s. The weak
$\Delta\omega$-dependence in the cavity model means that the
filtering effect is improved. On the other hand, if the atoms are
used to probe the state of the cavity field, then it may be
preferable to have a large $\Delta\omega$-dependence, see
\cite{qnd}.  We also see that, in the $l=2$ case, this does not
nescessarely imply that $|a_-^{\infty}(n)|^2=1$, even though the
area $A$ is zero. It is also clear from the figure that the $l=1$
and $l=3$ curves differs in this case.
%\begin{center}
%\vspace{1cm} Insert figure~\ref{fig3} about here \vspace{1cm}
%\end{center}

From the plots of the filter functions and the discussion above,
we conclude that it is possible to create a sharpened photon
distribution by letting atoms pass the microwave cavity. To check
this, we assume the initial photon distribution in the cavity to
be a Poissonian
\begin{equation}\label{costat}
|c_n|^2=\exp(-\bar{n})\frac{\bar{n}^n}{n!},
\end{equation}
centered around the second maximum of $|a_-^{\infty}(n)|^2$ at
$n=16$. If we measure $m$ atoms in their lower state after the
interactions, the photon distribution will be given by
(\ref{mner}). The distribution $P_n^-(m)$ is shown in figure 3
with $l=1$ and the other parameters are as in figure 1, the number
of atoms is $m=1$, $5$ and $25$ and initially $\bar{n}=16$. It is
manifest that the width of the distribution is decreasing when $m$
is increased.

Figure 3 shows clearly that a sharpened distribution is achived.
Another way to investigate the state of the field is to study the
Mandel $Q$-parameter, which is defined as
\begin{equation}
Q=\frac{\langle n^2\rangle-\langle n\rangle^2-\langle
  n\rangle}{\langle n\rangle}.
\end{equation}
For a Poissonian distribution, the $Q$-parameter is zero, for a
super-Poissonian state it is greater than zero and for a
non-classical sub-Poissonian state we have $-1<Q<0$. In figure 4
we have plotted the $Q$-parameter as a function of the
coupling-strength $g_0$ when twenty-five ($m=25$) atoms have
passed the cavity, and all of them have been recorded in their
lower state. The dimensionless parameters are $\Delta\omega=0.5$,
$T=0.1$ and $l=1$ and the initial field was in a coherent state
(\ref{costat}) with $\bar{n}=20$. The plot indicates that a
non-classical state is achived for a large range of values of
$g_0$. If the initial state of the field is known, the
$Q$-parameter can be used to find the optimal parameters $g_0$ and
$T$ needed to get a sharpened photon distribution.
%\begin{center}
%\vspace{1cm} Insert figure~\ref{fig4} about here \vspace{1cm}
%\end{center}

\section{Conclusion}\label{con}
In this paper we have discussed the
filtering action of a model where the time dependence of the coupling
constant is provided by the field experienced by an atom traversing a
cavity eigenmode. Numerically we have investigated a trigonometric
dependence on the position in the cavity. This model neglects possible
complications arising from edge effects at the entrance or exit of the
cavity. The resulting effect is expected to be small, but when the
situation is such that the onset of the interaction switches between sudden
and adiabatic, there are interesting questions to explore. This
was, in fact, investigated in the case of zero detuning in \cite{joshi}.

For the method presented above to work, it is crucial that the
atom-field interaction-time is considerably smaller than the
cavity loss time $\tau_{loss}$. Typical experimental values are
$v\sim 300$ m/s and $\nu\sim 50$ GHz, which gives the interaction
time
\begin{equation}
\tau_{int}=\frac{\pi}{kv}\sim 10 \mu s.
\end{equation}
For a high-$Q$ microwave cavity one can achieve cavity loss times
as large as $\tau_{loss}\sim 0.3$ s, from which we conclude that
it is possible to manipulate the field before it decays. Suitable
choice of the atomic transition makes the atomic decay times
sufficiently large too.

The trigonometric mode shape has also been investigated by Schlicher
\cite{schlicher}, but he considers mainly the ability of the atom to
follow the changing field adiabatically. He introduces no explicit
measurement on the atom and does not discuss the change of the cavity
field effected by the interaction with the atom.

In our work we find that the Rosen-Zener model is much more sensitive to
the detuning between the atoms and the radiation than the simple
trigonometric model. This may be understood to derive from the different
interaction times; in the trigonometric model, the total interaction time
is determined by the period of the mode function. In the Rosen-Zener model, on the
other hand, the interaction acts over an infinite interval. Because the
detuning defines the rate of oscillation of the two levels, the effect of
it may show up more dramatically when it acts over the longer time
interval. A smooth switch on/off will enhance the possibility
for the atom to follow the field.

In order to verify this conjecture, we have investigated a variety
of pulse shapes with different sharpness of switching on and off
the interaction. The trigonometric function disappears linearly
and the Rosen-Zener model exponentially. Faster switching is
achived by a square-wave pulse or a Gaussian. A slower switching
can be achived by using a Lorentzian. We thus have the switch on
and off rate occurring in the ordered sequence
\begin{equation}\label{sensdet}
\mathrm{square-wave}>\mathrm{trigonometric}>\mathrm{Gaussian}>\mathrm{Rosen-Zener}>\mathrm{Lorentzian}.
\end{equation}
In order to check the dependence on the detuning $\Delta\omega$
for these cases, we choose the minimum at $n=49$ in figure 2. The
shift of the value at this minimum is shown in figure 5. As we can
see the sensitivity is indeed given by the sequence in
(\ref{sensdet}). In the comparison the pulse area has been choosen
the same in each case; the other details of the pulses are given
in the Appendix.

The two-level problem with Lorentzian and Gaussian pulse shapes has been investigated before \cite{bambini} and \cite{robinson}. However, they did not discuss the dependence on the detuning or how the pulse shape affect the possibility for adiabatic following. As the models in their works are semiclassical, the photon statistics discussed here is, of course, not relevant in their papers.
%
%\begin{center}
%\vspace{1cm} Insert figure~\ref{fig5} about here \vspace{1cm}
%\end{center}

\section*{Appendix}
The pulse area is choosen to be
\begin{equation}
A=\int_{-\infty}^{\infty}g(t)dt=\pi Tg_0=\frac{g_0\pi}{10},
\end{equation}
so that setting the time-scale $T=0.1$ in the Rosen-Zener model
determines the normalization for the different cases. Table 1
gives the different shapes of the pulses and parameters used in
figure 5.

%\begin{center}
%\vspace{1cm} Insert table~\ref{tab1} about here \vspace{1cm}
%\end{center}
%
\begin{center}
\begin{tabular}{|c|c|}
\hline & \\
Pulse shape & Parameters \\ & \\
\hline & \\
Square-wave:
$g(t)=g_0\frac{1}{2}\left[\tanh\left(\frac{t+\tau}{t_s}\right)-\tanh\left(\frac{t-\tau}{t_s}\right)\right]$
    & $\tau=\pi T$, $t_s=0.02$ \\ & \\ \hline & \\ Trigonometric:
    $g(t)=g_0\cos(kv\,t)$ & $kv=\frac{2}{\pi T}$ \\ & \\ \hline & \\ Gaussian:
    $g(t)=\frac{g_0}{\sqrt{\pi}2\sigma}\exp\left[-\frac{(t/\tau)^2}{4\sigma^2}\right]$ & $\tau=\pi T$, $\sigma=0.3$ \\ & \\ \hline & \\ Rosen-Zener: $g(t)=g_0\mathrm{sech}\left(\frac{t}{T}\right)$ & $T=0.1$ \\ & \\ \hline & \\ Lorentzian: $g(t)=g_0\frac{(\gamma/\pi)^2}{(t/\tau)^2+\gamma^2}$ & $\tau=\pi T$, $\gamma=0.3$ \\ & \\ \hline
\end{tabular}
\\
\end{center}

The parameter $\tau=\pi T$ gives the chosen pulse area and $t_s$,
$\sigma$ and $\gamma$ are measures of the different
widths/steepnesses in the models. The widths are choosen such that
$g(0)\approx 1$ for all pulses. The coupling strength is as in
figure 2, $g_0=5$, and is the same for all models. Note that we
have used hyperbolic tangent functions for simulation of the
square-wave pulse. However, in the limit that $t_s$ goes to zero,
we obtain a step-function pulse which is analytically solvable.

%\newpage

%\include{captions}
%\include{tab}

\newpage

\begin{thebibliography}{12}

\bibitem{jaynes-cummings} Jaynes, E. T.  and Cummings, F. W., 1963, Proc. IEEE
\textbf{51}, 89.

\bibitem{shore-knight} Shore, B. W. and Knight, P. L., 1993,
 \textit{Journal of Modern Optics} \textbf{40}, 1195.

\bibitem{jonas} Larson, J. and Stenholm, S., \textit{Journal of Modern Optics}, to be published.

\bibitem{fredrik} Mattinson, F., Kira, M. and Stenholm, S., 2001, \textit{Journal of Modern Optics},
\textbf{48}, 889.

\bibitem{walther} Raithel, G., Wagner, C., Walther, H., Narducci, L. M. and Scully, M. O.. 1994,  \textit{Cavity Quantum Electrodynamics, Advances in Atomic,
molecular and Optical Physics, Supplement 2, }, edited by P. R. Berman
(Academic Press), pp 57-122.

\bibitem{field-prep} For preperation of Fock states see; Brattke, S., Varcoe, B. T. H., and Walther, H.,
 2000, \textit{Phys. Rev. Lett.}, \textbf{86}, 3534. The realization of EPR states between two cavity   modes is
  presented in; Raushenbeutel, A., Bertet, P., Osnaghi, S., Nogues, G.,
  Brune, M., Raimond, J. M. and Haroche, S., 2001, \textit{Phys. Rev. A}, \textbf{64},
  050301. The decoherence of Schr\"odinger cat states has been
  measured and is described in; Brune, M., Hagley, E., Dreyer, J., Maitre, X.,
  Maali, A., Wunderlich, C., Raimond, J. M. and Haroche, S., 1996,
  \textit{Phys. Rev. Lett.}, \textbf{77}, 4887.

\bibitem{schlicher} Schlicher, R. R., \textit{Optics Comm.}, 1988, \textbf{70}, 97.

\bibitem{tidmod} Fang, M. F., 1998, \textit{Physica A}, \textbf{259}, 193.

\bibitem{joshi} Joshi, A. and Lawande, S. V., 1993, \textit{Phys. Rev. A}, \textbf{48},
  2276.

\bibitem{qnd} Brune, M., Haroche, S., Lefevre, V., Raimond, J. M. and
  Zagury, N., 1990, \textit{Phys. Rev. Lett.}, 1990,  \textbf{65}, 976.

\bibitem{bambini} Bambini, A. and Lindberg, M., 1984, \textit{Phys. Rev. A}, \textbf{30}, 794.

\bibitem{robinson} Robinson, E. J., 1985, \textit{J. Phys. B}, \textbf{18}, 3687.

\end{thebibliography}
\end{document}